\documentclass[12pt]{article}
\usepackage{a4wide,graphicx,amssymb}

\newcommand{\z}{&\hspace*{-8pt}}

\begin{document}
\title{\vskip-3cm{\baselineskip14pt
\centerline{\normalsize\hfill TTP 04-04}
\centerline{\normalsize\hfill SFB/CPP-04-06}
\centerline{\normalsize\hfill hep-ph/0403026}
\centerline{\normalsize\hfill February 2004}
}
\vskip.4cm 
Pole- versus ${\rm MS}$-mass definitions\\in the electroweak theory
\vskip.4cm
}
\author{
M.~Faisst, J.~H.~K\"uhn, and O.~Veretin\\[0.5cm]
{\it Institut f\"ur Theoretische Teilchenphysik,} \\
{\it Universit\"at Karlsruhe, D-76128 Karlsruhe, Germany}
}
\date{}
\maketitle

\begin{abstract}
Two different two-loop relations between the pole- and the
$\overline{\rm MS}$-mass of the top quark have been derived in the
literature which were based on different treatments of the tadpole
diagrams. In addition, the limit $M_W^2/m_t^2\to0$ was employed in one
of the calculations. It is shown that, after appropriate
transformations, the results of the two calculations are in perfect
agreement. Furthermore we demonstrate that the inclusion of the
non-vanishing mass of the $W$-boson leads to small modifications only.
\end{abstract}

\section*{}

The so-called $\rho$-parameter, originally introduced in
\cite{veltman1}, plays an important role in precision tests of the
Standard Model. The dominant contribution from virtual bottom and top
quarks, $\Delta\rho_t$, is of order $G_Fm_t^2$ and was originally
evaluated in \cite{veltman2}. During the years, and with increasing
experimental precision, the calculation of $\Delta\rho_t$ has in a first
step been pushed to two-loops, including QCD effects of order $\alpha_s
G_Fm_t^2$ \cite{Djouadi} and purely electroweak corrections of order
$(G_Fm_t^2)^2$ \cite{2loopEW}. In a next step, the three-loop QCD
corrections were evaluated in \cite{Tarasov,CKS}. Recently the two
remaining three-loop contributions, of order $\alpha_s (G_Fm_t^2)^2$ and
$(G_Fm_t^2)^3$, were evaluated. The approximation $m_t^2\gg M_{W,Z}^2$
was employed, corresponding to the ``gaugeless'' limit of the
electroweak theory or, in other words, to a spontaneously broken Yukawa
theory.  In \cite{Faisst:2003px}  the mass of the Higgs boson was kept
as an independent parameter. Together with the results of
\cite{vanderBij:2000cg}, where the special case $M_H=0$ was considered,
this completes the prediction for $\Delta\rho_t$ in three-loop
approximation.

In \cite{Faisst:2003px,vanderBij:2000cg} $\Delta\rho_t$ was first
evaluated in the $\overline{\rm MS}$ scheme. This reduces the problem to
the calculation of vacuum diagrams which were evaluated with the
help of the computer-algebra programs MATAD \cite{MATAD} and EXP
\cite{EXP}. In a second step, the $\overline{\rm MS}$-result was
transformed to the on-shell scheme using the $\overline{\rm MS}$ to
on-shell relations of the top quark mass of order $\alpha_s G_Fm_t^2$
and $(G_Fm_t^2)^2$ respectively for the two problems of interest.
This relation is available in analytic form for the special cases
$M_H=0$ \cite{vanderBij:2000cg} and $M_H=m_t$  \cite{Faisst:2003px}.
For the generic case, with arbitrary $M_H$, it was obtained by employing
suitable expansions around the point $M_H=m_t$ and in the limit of large
Higgs mass.

Recently an independent two-loop calculation of the $\alpha_s G_Fm_t^2$
relation between pole- and $\overline{\rm MS}$-mass in the framework of
the full electroweak theory was presented \cite{kalmykov1} in
closed analytical form  for arbitrary Higgs- and non-vanishing $W$-mass.
This constitutes an important ingredient for many three-loop calculations
of order $O(\alpha^2\alpha_s)$, where the validity of the approximation
$M_W^2\ll m_t^2$ is doubtful. Furthermore it provides an independent
check of the corresponding relation obtained in \cite{Faisst:2003px}
with the help of expansion methods. The special case $M_W\to 0$ was
subsequently given in \cite{kalmykov2}. The purpose of this brief note
is to clarify the relation between the two seemingly different results.

\begin{figure}[th]
  \begin{center}
    \includegraphics[clip,width=9cm]{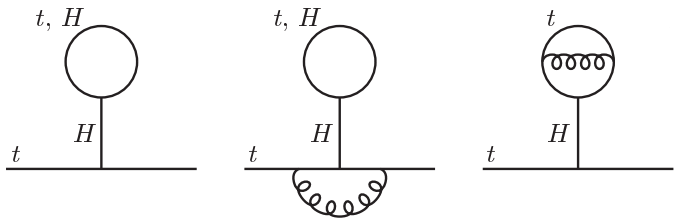}
  \end{center}
  \caption{\label{higgstad}One- and two-loop tadpole contributions to
    the self-energy.} 
\end{figure}

The renormalized self-energy  of a massive fermion with pole mass $M$ in
the on-shell scheme at one-loop order can be written as (we ignore
complications arising from the Dirac structures involving $\gamma_5$)
\begin{equation}
\Sigma_R^{\rm OS}( p ) = \Sigma_0( p ) - \Sigma_0( p ) \left. \right|_{
  p\!\!\!/=M} - M^2 \frac{\partial \Sigma_0(p)}{\partial p^2}
\left. \right|_{p\!\!\!/=M} \;,
\label{OS}
\end{equation}
where $\Sigma_0$ is the bare self-energy. One immediately finds that all
momentum independent contributions, in particular those from
tadpole-diagrams, cancel by construction. The same cancellation of
momentum independent terms occurs at the two-loop level, where in
(\ref{OS}) also mixed products of one-loop contributions have to be
considered.

In the $\rm MS$ schemes one just subtracts the singular part of the
Laurent expansion in $\varepsilon=(d-4)/2$ plus possibly some constant
term specific for the scheme. In this case we have 
\begin{equation}
\Sigma_R^{\rm MS}( p ) = \Sigma_0(p) - \Sigma_0(p)_{\rm div+const}\;. 
\label{MS}
\end{equation}
As a consequence, the prescription how to subtract constant terms
does affect the definition of the {\rm MS}-mass. One such constant
contribution to $\Sigma_0$ arises from the Higgs tadpole diagrams
(see Fig. \ref{higgstad}). In \cite{kalmykov1,kalmykov2} these
tadpole diagrams were included in the definition of the ${\rm MS}$-mass
and their contribution remains present in the final result for the ${\rm
  MS}$-pole-mass relation. In contrast, throughout the calculation in
\cite{Faisst:2003px,vanderBij:2000cg} the vanishing of the Higgs tadpole
was used as one of the renormalization conditions (see
e.g. \cite{denner} Eq. (3.4)). Therefore these tadpoles were absent in
the definition of the ${\rm MS}$-mass and, correspondingly, in the
evaluation of the diagrams relevant for the $\rho$-parameter, as
required for a consistent result. (For early discussions of this issue
at the one- and two-loop level see e.g. \cite{discussion:tad})

Since the strategy for the evaluation of the Feynman amplitudes is
entirely different in \cite{kalmykov1,kalmykov2} compared to
\cite{Faisst:2003px,vanderBij:2000cg} (expansions vs. closed analytic
formulae), a comparison between the two results seems desirable. We
therefore include the Higgs tadpole diagrams in the calculation of the
${\rm MS}$ top quark mass $m_{t,\rm tadp}$ based on
\cite{Faisst:2003px,vanderBij:2000cg}. The impact of the tadpole
diagrams shown in Fig. \ref{higgstad} is given by the ratio between
$m_{t,\rm tadp}$ and $m_{t,\rm notadp}$ calculated for
\cite{Faisst:2003px,vanderBij:2000cg}. In order $O(\alpha_s G_F^2m_t^4)$
it reads 
\begin{eqnarray}
\frac{m_{t,\rm tadp}(\mu)}{m_{t,\rm notadp}(\mu)} \z=\z 
   1 + X_t \left( 
         - \frac32 \frac{M_H^2}{M_t^2} 
         + 4 N_c \frac{M_t^2}{M_H^2}
         + \frac32 \frac{M_H^2}{M_t^2} \log\frac{M_H^2}{\mu^2} 
         - 4 N_c \frac{M_t^2}{M_H^2} \log\frac{M_t^2}{\mu^2} 
        \right) \nonumber\\
\z\z
     + C_F \frac{\alpha_s}{4\pi} X_t \left(
           8 N_c \frac{M_t^2}{M_H^2}
         \!+\! 48 N_c \frac{M_t^2}{M_H^2} \log\frac{M_t^2}{\mu^2} 
         \!-\! 24 N_c \frac{M_t^2}{M_H^2} \log^2\frac{M_t^2}{\mu^2} 
         \right),
\label{m2m}
\end{eqnarray}
where $M_H$ and $M_t$ are pole (on-shell) masses and the gaugeless limit
has been employed.
\begin{figure}[b]
\begin{center}
\includegraphics[clip,width=10cm]{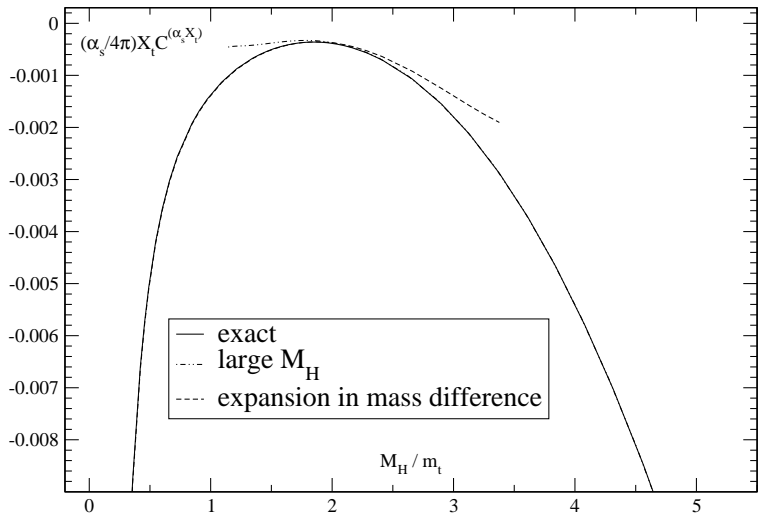}
\end{center}
\caption{\label{comparision}
  The two-loop coefficient $C^{(\alpha_s X_t)}$ including the tadpole terms.
  The solid line represents the analytical result,
  the dashed line the expansion in the large $M_H$ limit, the dash-dotted line 
  the expansion around $M_H=M_t$. All expansions are performed to fifth
  order.}
\end{figure}

Equation (\ref{m2m}) can now be used to compare the two-loop relations
between the $\overline{\rm MS}$ and the pole mass based on
\cite{kalmykov1,kalmykov2} and \cite{Faisst:2003px,vanderBij:2000cg}
respectively. This relation can be written as
\begin{equation}
  \frac{m(\mu)}{M} = 1 + \frac{\alpha_s}{\pi} C^{(\alpha_s)}
         + X_t C^{(X_t)} + \frac{\alpha_s}{\pi}X_t C^{(\alpha_s X_t)} 
         + \dots, 
\end{equation}
\begin{equation}
   X_t = \frac{G_F M_t^2}{8\sqrt{2}\pi^2} \approx 3\times 10^{-3} \,.
\end{equation}
The coefficients $C^{(X_t)}$ and $C^{(\alpha_s X_t)}$ depend on the
prescription. In the gaugeless limit they are functions of $M_H^2/M_t^2$
only.

\begin{figure}[t]
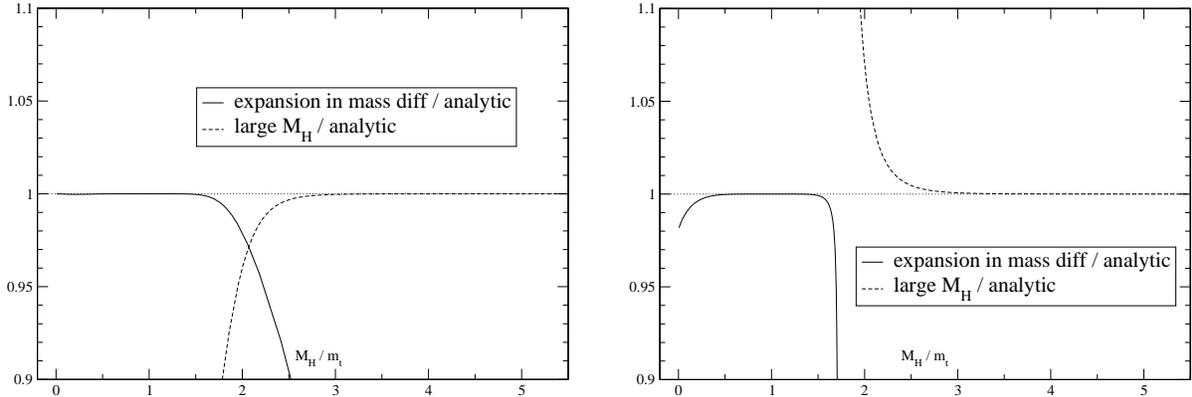

\begin{center}
\includegraphics[clip,width=7.5cm]{plot2.eps} \hspace{0.5cm}
\includegraphics[clip,width=7.5cm]{plot2a.eps}
\end{center}
\caption{\label{ratios}
  Ratio between expanded and analytic results in the scheme with (left
  figure) and without (right figure) tadpoles. The dashed line
  corresponds to the expansion in the large $M_H$ limit, the solid
  line to the expansion around $M_H=M_t$.}
\end{figure}

The result for the tadpole terms separately exhibits a power law
behaviour in the limit $M_H^2/M_t^2\to0$ and the limit $M_t^2/M_H^2\to0$
whereas the complete result without tadpoles remains finite for
$M_H^2/M_t^2\to0$. Using Eq. (\ref{m2m}) we find agreement between the
results of \cite{Faisst:2003px,vanderBij:2000cg} and
\cite{kalmykov1,kalmykov2}. This is demonstrated in
Fig.~\ref{comparision} where we present the results for the two-loop 
coefficient $C^{(\alpha_s X_t)}$ in the gaugeless limit employing the
definition which includes the tadpole terms.

The corresponding ratios between the expanded and the analytic results
are shown in Fig.~\ref{ratios} for both schemes (with and without
tadpoles). From this comparison it is evident that
the two calculations \cite{Faisst:2003px,vanderBij:2000cg} and
\cite{kalmykov1,kalmykov2} do agree for the relation between pole- and 
$\overline{\rm MS}$-mass after compensating for the tadpole
contributions, and that the expansion with five terms give an excellent
approximation to the analytic result with less than 10\% deviation at
most and negligible deviation for the physically interesting range of
the Higgs mass. In particular the agreement between the expansion around
$M_H=M_t$ and the analytic result for small $M_H$ is remarkable, as
already observed in \cite{Faisst:2003px}.

\begin{figure}[ht]
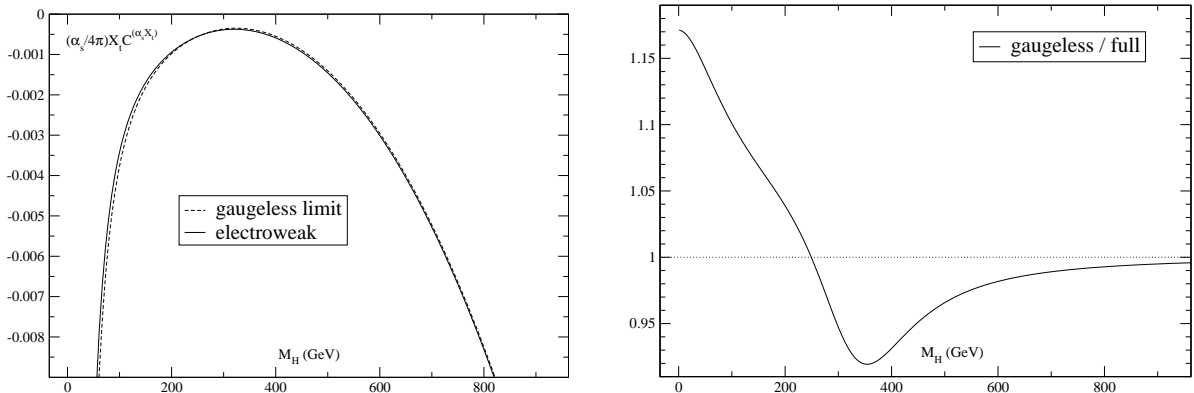

\begin{center}
\includegraphics[clip,width=7.5cm]{plot3.eps} \hspace{0.5cm}
\includegraphics[clip,width=7.5cm]{plot4.eps}
\end{center}
\caption{\label{full}
  Comparison between $C^{(\alpha_s X_t)}$ evaluated in the full
  electroweak theory and the gaugeless approximation. In the left figure,
  the solid line represents full electroweak theory, the dashed line the
  gaugeless limit. In the right figure, the ratio between the results in
  the full electroweak theory and the gaugeless limit for the two-loop
  coefficient $C^{(\alpha_s X_t)}$ is shown.
}
\end{figure}

It is also instructive to compare the result obtained in the gaugeless
limit  with the one \cite{kalmykov1} obtained in the full electroweak
theory with nonvanishing $M_W$. In \cite{kalmykov2} it was shown that
this difference is given by the following expression
\begin{eqnarray}
&& 
\frac{\overline{m}^{\rm SM}_{t,\rm tadp}(M_t) - \overline{m}^{\rm
    g.l.}_{t,\rm tadp}(M_t)}{M_t~X_t }  
 = 
 - 0.07978 
 - 0.429164  \frac{\alpha_s}{4 \pi} C_f
\nonumber \\ && \qquad \qquad 
+\frac{M_t^2}{M_H^2} \left( 1 - 4 \frac{\alpha_s}{4 \pi} C_f \right)
\Biggl[
- \frac{1}{2} \frac{M_W^4}{M_t^4} \left( 1 - 3 \ln \frac{M_W^2}{M_t^2} \right)
- \frac{1}{4} \frac{M_Z^4}{M_t^4} \left( 1 - 3 \ln \frac{M_Z^2}{M_t^2} \right)
\Biggr]
\;. 
\nonumber 
\end{eqnarray} 
The two-loop coefficients are compared in Fig.~\ref{full}. The deviation
is small and does not exceed 10\% for $M_H>100$GeV.

Summary: The difference between \cite{Faisst:2003px,vanderBij:2000cg}
and \cite{kalmykov1,kalmykov2} results from the exclusion of tadpole
diagrams, which do not contribute to physical observables in the
on-shell scheme. The results based on expansions around $M_H=M_t$ and
the limit of large $M_H$ \cite{Faisst:2003px} are in perfect numerical
agreement with the analytic results \cite{kalmykov1,kalmykov2}. The
influence of non-zero $M_W$-mass terms is below 10\% for $M_H>100$GeV,
the region of interest for phenomenology.

{\it Acknowledgments:} 
The authors would like to thank K.~Chetyrkin, M.~Yu.~Kalmykov,
D.~Kazakov, and F.~Jegerlehner for helpful discussions and
M.~Yu.~Kalmykov for providing the data in the full electroweak theory.

  This work was supported by the {\it Graduiertenkolleg
``Hochenergiephysik und Teil\-chen\-astrophysik''}, by BMBF under grant
No. 05HT9VKB0, and the SFB/TR 9 (Computational Particle Physics).


\begin{thebibliography}{99}

\bibitem{veltman1}
D.~A.~Ross and M.~J.~G.~Veltman,
Nucl.\ Phys.\ B {\bf 95} (1975) 135.

\bibitem{veltman2}
M.~J.~G.~Veltman,
Nucl.\ Phys.\ B {\bf 123} (1977) 89.


\bibitem{Djouadi}
A.~Djouadi and C.~Verzegnassi,
Phys.\ Lett.\ B {\bf 195} (1987) 265;\\
A.~Djouadi,
Nuovo Cim.\ A {\bf 100} (1988) 357.

\bibitem{2loopEW}
J.~J.~van der Bij and F.~Hoogeveen,
Nucl.\ Phys.\ B {\bf 283} (1987) 477;\\
J.~Fleischer, O.~V.~Tarasov, and F.~Jegerlehner,
Phys.\ Lett.\ B {\bf 319} (1993) 249.

\bibitem{Tarasov}
L.~Avdeev et al., 
Phys.\ Lett.\ B {\bf 336} (1994) 560;
[Erratum-ibid.\ B {\bf 349} (1995) 597].

\bibitem{CKS}
K.~G.~Chetyrkin, J.~H.~Kuhn, and M.~Steinhauser,
Phys.\ Lett.\ B {\bf 351} (1995) 331.


\bibitem{Faisst:2003px}
M.~Faisst et al.,
Nucl.\ Phys.\ B {\bf 665} (2003) 649.

\bibitem{vanderBij:2000cg}
J.~J.~van der Bij et al., 
Phys.\ Lett.\ B {\bf 498} (2001) 156.

\bibitem{MATAD}
M.~Steinhauser,
Comput.\ Phys.\ Commun.\  {\bf 134} (2001) 335.

\bibitem{EXP}
T. Seidensticker, Diploma thesis (University of Karlsruhe, 1998), unpublished.

\bibitem{kalmykov1}
F.~Jegerlehner and M.~Y.~Kalmykov, 
Nucl.Phys. B {\bf 676} (2004) 365.

\bibitem{kalmykov2}
F.~Jegerlehner and M.~Y.~Kalmykov,
Acta Phys.\ Pol.\ B {\bf 34} (2003) 5335.

\bibitem{denner}
A.~Denner,
Fortsch.\ Phys.\  {\bf 41} (1993) 307.

\bibitem{discussion:tad}
T.~Appelquist et al.,
Phys.\ Rev.\ D {\bf 8} (1973) 1747;


J.~Fleischer and F.~Jegerlehner,
Phys.\ Rev.\ D {\bf 23} (1981) 2001;

R.~Hempfling and B.~A.~Kniehl,
Phys.\ Rev.\ D {\bf 51} (1995) 1386.

\end{thebibliography}
\end{document}